\documentclass[11pt]{article}

\usepackage[margin=1in]{geometry}
\usepackage{amsmath,amssymb}
\usepackage{booktabs}
\usepackage{graphicx}
\usepackage{hyperref}
\usepackage[T1]{fontenc}
\usepackage[utf8]{inputenc}
\usepackage{lmodern}
\usepackage{microtype}
\usepackage{array}
\usepackage{listings}
\usepackage{xcolor}
\usepackage{authblk}
\usepackage{natbib}
\setcitestyle{authoryear,round}

\hypersetup{
  colorlinks=true,
  linkcolor=blue,
  citecolor=blue,
  urlcolor=blue,
  pdfcreator={LaTeX}
}

\title{Symbolic-Vector Attention Fusion for Collective Intelligence}

\author{Hongwei Xu}
\affil{SYM.BOT \\ \texttt{hongwei@sym.bot}}

\date{April 2026}

\begin{document}
\maketitle

\begin{abstract}
When autonomous agents observe different domains of a shared environment, each signal they exchange mixes relevant and irrelevant dimensions. No existing mechanism lets the receiver evaluate \emph{which} dimensions to absorb. We introduce \textbf{Symbolic-Vector Attention Fusion (SVAF)}, the content-evaluation half of a two-level coupling engine for collective intelligence. SVAF decomposes each inter-agent signal into 7 typed semantic fields, evaluates each through a learned fusion gate, and produces a \emph{remix} --- new knowledge from the intersection of two domains. A band-pass model yields four outcomes (redundant, aligned, guarded, rejected), solving both selectivity and redundancy. The fusion gate independently discovers a cross-domain relevance hierarchy: mood emerges as the highest-weight field by epoch~1, before accuracy plateaus --- consistent with independent mechanistic evidence that LLM emotion representations are structurally embedded along valence--arousal axes. SVAF forms Layer~4 of the Mesh Memory Protocol (MMP); the other half of the coupling engine is a per-agent Closed-form Continuous-time (CfC) neural network at Layer~6, whose learned per-neuron time constants ($\tau$) create the temporal dynamics from which collective intelligence emerges: fast neurons synchronise affect across agents in seconds, while slow neurons preserve domain expertise indefinitely. SVAF determines \emph{what} enters each agent's cognitive state; CfC determines \emph{how} that state evolves. Trained on 237K samples from 273 narrative scenarios, SVAF achieves 78.7\% three-class accuracy. We verify the complete mesh cognition loop --- from per-field evaluation through remix, CfC state evolution, $\tau$-modulated peer blending, and autonomous action --- in a live deployment with 7 nodes across macOS, iOS, and web.
\end{abstract}

\section{Introduction}

Multi-agent systems increasingly deploy heterogeneous agents that observe different aspects of a shared environment. A coding assistant monitors developer activity; a music agent tracks listening patterns; a fitness agent detects sedentary behaviour. When these agents operate in isolation, each holds partial understanding. The fundamental question is: \textbf{how should an agent on the mesh evaluate and absorb observations from peers in different domains, when some dimensions of the incoming signal are relevant and others are not?}

This problem is distinct from multi-agent communication \citep{sukhbaatar2016learning,das2019tarmac}, which learns what to send; from agent memory \citep{park2023generative,packer2023memgpt}, which stores and retrieves within a single agent; and from multi-agent debate \citep{du2024improving}, which reconciles disagreements on a shared question. The problem we address is \textbf{cross-domain signal evaluation}: given a structured signal from an agent in a different domain, determine per-field which dimensions to accept, which to suppress, and how to fuse the accepted fields with local memory to produce new understanding.

No existing work addresses this problem directly. The closest analogue is single-vector cosine similarity, used implicitly in agent memory systems \citep{park2023generative,packer2023memgpt} for retrieval and in multi-agent communication \citep{sukhbaatar2016learning} for signal routing. In all cases, the entire signal is encoded into one vector and compared via a single distance score. This default approach fails when relevant and irrelevant dimensions coexist in the same signal, and cannot distinguish redundancy from novelty (\S\ref{sec:scalar-limitation}).

We introduce \textbf{Symbolic-Vector Attention Fusion (SVAF)}, a neural per-field memory evaluation and fusion mechanism for multi-agent systems. SVAF operates within the Mesh Memory Protocol (MMP; \citealt{xu2026mmp}), a mesh protocol for collective intelligence. In MMP's 8-layer architecture, SVAF serves as the Layer~4 evaluation gate --- it determines what enters each agent's cognitive state and how it is absorbed. The protocol layers below (identity, transport, connection, memory) handle peer discovery and frame delivery; the layers above (synthetic memory, per-agent neural networks, application) handle reasoning and action. SVAF is the boundary between protocol infrastructure and mesh cognition.

SVAF evaluates signals structured as \textbf{Cognitive Memory Blocks (CMBs)} --- each observation decomposed into 7 typed semantic fields (the CAT7 schema). The receiving agent's SVAF evaluates each field independently through a learned fusion gate, conditioned on cross-field attention, per-agent temporal freshness, and sender confidence. Accepted fields are fused with local memory to produce a \textbf{remix} --- not a copy, but new knowledge reflecting both the incoming signal and the receiver's domain context.

SVAF is one half of the coupling engine. The other half is a per-agent \textbf{Closed-form Continuous-time (CfC) neural network} \citep{hasani2022closed} at Layer~6, whose learned per-neuron time constants ($\tau$) create the temporal dynamics from which collective intelligence emerges. Fast neurons ($\tau < 5$s) synchronise mood across agents in seconds; slow neurons ($\tau > 30$s) preserve domain expertise indefinitely. SVAF determines \emph{what} enters each agent's cognitive state; CfC determines \emph{how} that state evolves. Together they solve the dual problem of collective intelligence: sharing awareness without losing individual expertise.

MMP implements a two-level coupling architecture: peer-level drift (are we cognitively close?) and content-level SVAF (is this specific signal useful?). These operate independently --- a rejected peer can send highly relevant CMBs --- enabling cold-start bootstrapping between agents with no shared history (\S\ref{sec:live-deployment}).

This paper makes four contributions:

\textbf{C1: Structured decomposition for cross-domain evaluation.} Cognitive Memory Blocks (CMBs) with 7 typed semantic fields (CAT7) --- a fixed, shared schema that enables any two agents to evaluate each other's signals. (\S\ref{sec:cmb})

\textbf{C2: Per-field evaluation and remix.} A neural fusion gate that evaluates each field independently and produces a remix --- new knowledge from domain intersection, not a copy. The growing DAG of remixes, processed by each agent's CfC neural network with $\tau$-modulated temporal dynamics, is the substrate from which collective intelligence emerges. (\S\ref{sec:method}, \S\ref{sec:remix})

\textbf{C3: Protocol-level context engineering for multi-LLM systems.} Each agent's LLM reasons on a projected subgraph curated by per-agent field weights and lineage tracing, with SVAF as the inter-LLM interface. Context quality --- what enters each LLM's window --- is determined by the protocol, not by application-layer heuristics. (\S\ref{sec:discussion})

\textbf{C4: Empirical cross-domain relevance hierarchy.} The fusion gate discovers mood as the universally relevant field by epoch~1, before accuracy plateaus --- consistent with affect being the most cross-domain-relevant dimension. (\S\ref{sec:gate-analysis})

\section{Related Work}
\label{sec:related-work}

\textbf{Multi-agent communication.} CommNet \citep{sukhbaatar2016learning} and TarMAC \citep{das2019tarmac} learn agent communication via mean-pooled or attention-targeted messages. These operate on whole hidden states in discrete time steps --- no per-field decomposition, no temporal decay, no receiver-side evaluation. Multi-agent debate \citep{du2024improving} reconciles disagreements between agents on the \emph{same} question; SVAF addresses a different problem --- evaluating heterogeneous observations from agents in \emph{different} domains. SVAF operates on structured memory fields in continuous time with receiver-autonomous evaluation.

\textbf{Agent memory.} Generative Agents \citep{park2023generative} maintain per-agent memory streams with retrieval. MemGPT \citep{packer2023memgpt} manages LLM memory as a virtual context hierarchy with self-directed retrieval. HippoRAG \citep{jimenez2024hipporag} uses knowledge-graph structure. These focus on single-agent memory; SVAF addresses multi-agent cross-domain fusion with drift-bounded evaluation.

\textbf{Multi-agent coordination.} \citet{shen2026empirical} empirically compare subagent and agent-team architectures for automated research, finding a trade-off between operational stability and theoretical deliberation. Their agents share a global memory file but lack per-field evaluation --- all agents see the same memory regardless of domain relevance. \citet{lyu2026evoscientist} introduce EvoScientist, where specialised agents maintain persistent memories for scientific discovery --- but their memory is centralised and domain-homogeneous. \citet{nguyen2026coordination} survey multi-agent LLM systems for financial trading and propose the \emph{Coordination Primacy Hypothesis}: inter-agent coordination protocol design is a primary driver of decision quality, often exerting greater influence than model scaling. This finding from a domain entirely distinct from ours provides independent corroboration of SVAF's premise --- that the protocol layer (how agents evaluate and fuse signals) matters more than the application layer (which LLM each agent runs). Their work identifies the problem; SVAF provides a concrete mechanism. \citet{bi2026cascade} address \emph{communication scope}: CASCADE introduces gate-triggered scoped propagation where agents expand their coordination footprint only when local validation gates indicate the current scope is insufficient. CASCADE controls \emph{who} receives a signal via scope budgets; SVAF controls \emph{what} a receiver absorbs via per-field drift thresholds. The two are complementary: CASCADE's scoped propagation could serve as transport-layer routing within MMP, while SVAF provides content-level evaluation that CASCADE's gates do not address.

\textbf{Agent interoperability protocols.} Model Context Protocol (MCP; \citealp{anthropic2024mcp}) standardises how LLMs access tools and data sources. Agent-to-Agent Protocol (A2A; \citealp{google2025a2a}) defines task delegation between agents via Agent Cards. Both address agent \emph{integration}; neither addresses cross-domain \emph{signal evaluation}. SVAF operates at a different layer --- evaluating the \emph{content} of inter-agent signals per-field, enabling selective absorption across domain boundaries.

\textbf{Affective representations in language models.} SVAF treats affect as a first-class architectural element: the mood field carries structured valence--arousal coordinates and receives a protocol-level guarantee (R5) --- even when SVAF rejects a CMB, the mood field is always delivered. Recent mechanistic interpretability work provides independent evidence: \citet{sofroniew2026emotion} demonstrate that emotion representations inside an LLM spontaneously organise along valence and arousal axes ($r = 0.81$, $r = 0.66$ against human ratings), reproducing Russell's circumplex model \citep{russell1980circumplex}. These vectors \emph{causally} influence model behaviour --- they are structurally embedded features, not surface sentiment correlates. This validates SVAF's design: the valence--arousal parameterisation aligns with LLM-internal geometry, and affect's causal role within a single model suggests it carries the most actionable information \emph{between} models in different domains. Our fusion gate independently confirms this hierarchy (\S\ref{sec:gate-analysis}).

\textbf{Selective aggregation.} Attention Residuals \citep{kimiteam2026attnres} replaces fixed residual connections in LLMs with learned softmax attention over preceding layer outputs, showing that uniform accumulation dilutes each layer's contribution. The analogy to multi-agent signals is direct: uniform field weighting (scalar evaluation) dilutes relevant fields with irrelevant ones. Where AttnRes learns selective aggregation over \emph{network depth}, SVAF learns it over \emph{agent signal fields}. Both replace a fixed all-ones mixing matrix with learned, content-dependent weights. SVAF's band-pass extension (\S\ref{sec:fusion}) adds a lower bound on novelty, rejecting signals too similar to existing state --- selectivity alone does not detect redundancy.

\section{Problem Formulation}

\subsection{Setting}

We consider $N$ autonomous agents $\{A_1, \ldots, A_N\}$ on an agent-to-agent mesh, each operating in a distinct domain. Agents produce structured observations about a shared environment --- a user's state, a warehouse floor, a research question, a patient's condition --- and broadcast them to all peers. Each receiving agent independently evaluates incoming signals and decides how to remix them with local memory, producing new understanding rather than storing copies. There is no central model, no orchestrator, and no shared memory store.

\subsection{Requirements}

\textbf{R1: Per-field selectivity.} Accept relevant fields of a cross-domain signal while suppressing irrelevant fields. A fitness agent receiving a coding agent's fatigue signal should absorb the mood (exhausted) and suppress the focus (code details).

\textbf{R2: Receiver autonomy.} Each agent evaluates based on its own domain priorities ($\alpha_f$ field weights). No central authority decides what is relevant --- the receiver decides.

\textbf{R3: Remix, not copy.} The stored result is a new CMB --- a remix reflecting both the incoming signal and local context, with lineage tracing back to its parents. The original is not stored.

\textbf{R4: Per-agent temporal awareness.} Stale signals are down-weighted via continuous temporal decay with agent-specific freshness windows ($\tau_i$). A music agent's 30-minute window and a fitness agent's 3-hour window reflect different domain temporal scales.

\textbf{R5: Cross-domain relevance.} Some fields (mood/affect) should couple across all domains. Others (perspective, intent) should remain sovereign to the receiver. The mechanism should discover --- or at minimum enforce --- this hierarchy.

\textbf{R6: Auditability.} Every remix event is reconstructible: which fields were accepted, with what gate values, from which parent CMBs, producing which lineage.

\textbf{R7: Redundancy detection.} Signals semantically identical to existing local memory must be rejected regardless of domain relevance. A paraphrase carries zero information gain (Shannon, 1948) and should not produce a remix. The mechanism must distinguish ``novel and relevant'' from ``redundant.''

\subsection{Why Single-Vector Similarity Is Insufficient}
\label{sec:scalar-limitation}

No prior work proposes a dedicated mechanism for cross-domain signal evaluation. The implicit default across agent memory \citep{park2023generative,packer2023memgpt} and multi-agent communication \citep{sukhbaatar2016learning,das2019tarmac} is single-vector cosine similarity: encode the entire signal into one vector, compute one distance $\delta = 1 - \cos(E(m), h_{\text{local}})$, and accept if $\delta \leq \theta$.

This is not a solution being critiqued --- it is the \emph{absence} of a solution. Single-vector similarity was not designed for cross-domain signal evaluation; it is a general-purpose retrieval primitive repurposed as a decision mechanism. It violates R1 (relevant and irrelevant dimensions conflated into one score), R5 (cannot discover per-field coupling hierarchies), and R7 (paraphrases produce different vectors, pass the threshold, and are stored as if novel).

\subsection{Two-Level Coupling Architecture}

The Mesh Memory Protocol addresses these requirements through a hierarchical coupling architecture with two independent evaluation levels:

\begin{table}[h]
\centering
\small
\begin{tabular}{@{}llll@{}}
\toprule
Level & Question answered & Operates on & Mechanism \\
\midrule
Peer-level & Is this agent cognitively close? & CfC hidden states & Cosine drift \\
Content-level (SVAF) & Is this signal relevant to me? & CMB fields & Per-field drift + fusion gate \\
\bottomrule
\end{tabular}
\end{table}

Peer-level coupling classifies the relationship between two agents:
\begin{itemize}
\item \textbf{Aligned} ($\delta \leq 0.25$): cognitive states are close; blending strength $\alpha = 0.40$
\item \textbf{Guarded} ($0.25 < \delta \leq 0.50$): moderate distance; $\alpha = 0.15$
\item \textbf{Rejected} ($\delta > 0.50$): cognitively distant; $\alpha = 0$
\end{itemize}

Content-level evaluation (SVAF) operates \textbf{independently} of the peer-level decision. Note that peer-level coupling has no redundancy zone: low peer drift means cognitive alignment, which is desirable. Low \emph{content} drift means the specific signal is already in memory, which is redundant. The asymmetry reflects the fundamental difference between coupling state (continuous, evolving) and signal content (discrete, potentially repeated). SVAF uses a \textbf{band-pass} model with four classes:
\begin{itemize}
\item \textbf{Redundant} ($\max_f(\delta_f) < T_{\text{redundant}}$; default 0.10): all fields already in memory; signal discarded
\item \textbf{Aligned} ($\delta_{\text{total}} \leq T_{\text{aligned}}$; default 0.25): novel and relevant; full fusion
\item \textbf{Guarded} ($T_{\text{aligned}} < \delta_{\text{total}} \leq T_{\text{guarded}}$; default 0.50): cautious acceptance; attenuated fusion
\item \textbf{Rejected} ($\delta_{\text{total}} > T_{\text{guarded}}$): irrelevant domain; discarded (mood still delivered, R5)
\end{itemize}
The redundancy test (R7) is per-field: a signal is redundant only if \emph{every} field falls below $T_{\text{redundant}}$. If any field carries novel content, the signal passes (\S\ref{sec:fusion}).

A rejected peer can send a highly relevant CMB --- SVAF evaluates the content, not the sender's overall cognitive proximity. This independence enables cold-start bootstrapping: when two agents first connect, peer drift is high (typically $> 0.8$), but SVAF can accept relevant CMBs immediately. Each accepted CMB shifts the receiver's cognitive state, narrowing peer drift over subsequent cycles (verified in production: 0.936~$\to$~0.468 in one cycle, \S\ref{sec:live-deployment}).

The mood field has a protocol-level guarantee (R5): when SVAF rejects a CMB, the mood field is still delivered if non-neutral. Affective state crosses all domain boundaries --- supported by independent evidence that LLM emotion representations are structurally embedded along valence--arousal axes (\S\ref{sec:related-work}).

SVAF is the content-level mechanism in this architecture. The remainder of this paper presents the CMB data structure (\S\ref{sec:cmb}), SVAF's design (\S\ref{sec:method}), evaluation (\S\ref{sec:evaluation}), and role in collective intelligence (\S\ref{sec:remix}).

\section{Cognitive Memory Blocks}
\label{sec:cmb}

Cognitive Memory Blocks (CMBs) were originally defined in the Mesh Memory Protocol \citep{xu2026mmp}. A CMB decomposes any observation into 7 structured semantic fields that can be independently evaluated across domain boundaries. This section explains why the schema exists, why it has exactly 7 fields, and why it is fixed rather than extensible.

\textbf{Why structured fields.} Scalar evaluation encodes an entire observation into one vector and computes one cosine distance. This loses the per-dimension nuance that SVAF requires. ``User exhausted after 8 hours coding with an AI assistant'' contains a mood (exhausted), an issue (fatigue), a focus (coding session), and a perspective (developer). A fitness agent needs the mood and issue but not the focus or perspective. Structured fields make this selectivity possible.

\textbf{Why 7 fields.} The 7 fields form a compact basis spanning three axes of human communication: \textbf{what} (focus, issue), \textbf{why} (intent, motivation, commitment), and \textbf{who/when/how} (perspective, mood). They are universal --- the same 7 fields serve every domain. Domain-specific interpretation happens in the field text, not the field name. A coding agent's \texttt{focus} is ``building auth module with AI assistant''; a fitness agent's \texttt{focus} is ``30-minute HIIT workout.'' Same field, different domain lens.

\textbf{Why a shared schema is required.} SVAF computes per-field drift across all 7 dimensions. If each agent defined its own fields, cross-domain evaluation would be impossible --- a fitness agent and a music agent would have no common dimensions to compute drift on. The fixed schema enables any two agents to evaluate each other's signals, regardless of domain. Fields the agent cannot meaningfully extract are set to ``neutral'' (a known, consistent baseline vector), never omitted --- this keeps the drift formula well-defined.

Formally, a CMB is a set of typed field-vector pairs:
\begin{equation}
\text{CMB} = \{(f, t_f, v_f) : f \in \mathcal{F}\}
\end{equation}
where $\mathcal{F}$ is the 7-element field schema, $t_f$ is a symbolic text label (human-readable for audit and retrieval), and $v_f \in \mathbb{R}^d$ is a unit-normalized vector embedding (machine-comparable for drift evaluation).

\begin{table}[h]
\centering
\small
\begin{tabular}{@{}llll@{}}
\toprule
Field $f$ & Axis & What it captures & Cross-Domain Relevance \\
\midrule
\texttt{focus} & Subject & What the text is centrally about & Medium \\
\texttt{issue} & Tension & Risks, gaps, assumptions, open questions & Medium \\
\texttt{intent} & Goal & Desired change or purpose & Low \\
\texttt{motivation} & Why & Reasons, drivers, incentives & Medium \\
\texttt{commitment} & Promise & Who will do what, by when & Medium \\
\texttt{perspective} & Vantage & Whose viewpoint, situational context & Low \\
\texttt{mood} & Affect & Emotion (valence $\in [-1,1]$, arousal $\in [-1,1]$) & \textbf{Fast} \\
\bottomrule
\end{tabular}
\caption{The CAT7 field schema. Each CMB contains exactly these 7 fields.}
\label{tab:cat7}
\end{table}

\textbf{Symbolic-vector dual representation.} Each field carries two representations: a \textbf{symbolic text label} $t_f$ (human-readable, used for audit, retrieval, and LLM reasoning) and a \textbf{unit-normalised vector embedding} $v_f \in \mathbb{R}^d$ (machine-comparable, used for drift computation and fusion). This duality is the ``Symbolic-Vector'' in SVAF. The text preserves meaning for agents that reason in natural language; the vector enables the mathematical operations (cosine similarity, weighted fusion, gated interpolation) that SVAF requires. Encoding is performed by a shared backbone with per-field projection heads (\S\ref{sec:encoder}): each field learns its own vector subspace, so ``tired'' in the mood field and ``tired'' in the focus field produce different embeddings tuned to their respective semantic axes.

\textbf{Why mood is special.} Mood is the only fast-coupling field --- affective state crosses all domain boundaries. The \texttt{mood} field carries numeric valence and arousal values based on Russell's circumplex model \citep{russell1980circumplex} --- the same two-dimensional structure that emerges inside LLMs (\S\ref{sec:related-work}). Other fields are inherently symbolic. The fusion gate independently confirms this design by discovering mood as the highest-weight field (\S\ref{sec:gate-analysis}).

\textbf{Concrete example.} A coding assistant observes fatigue and produces:

\begin{lstlisting}
CMB:
  focus:       "coding session with AI assistant, 8 hours"
  issue:       "exhausted, making simple mistakes"
  intent:      "needs a break before continuing"
  motivation:  "prevent bugs from fatigue-driven errors"
  commitment:  "coding session ongoing but degrading"
  perspective: "developer, afternoon, no breaks taken"
  mood:        "frustrated, low energy" (v: -0.6, a: -0.4)
\end{lstlisting}

A fitness agent evaluating this CMB via SVAF would strongly accept \texttt{mood} (frustration, low energy $\to$ recommend movement), accept \texttt{issue} (exhaustion $\to$ fitness-relevant), and suppress \texttt{perspective} (developer viewpoint $\to$ irrelevant to exercise). This per-field selectivity is what scalar evaluation cannot provide --- and it is the core problem SVAF solves.

New agent types join the mesh by defining their per-agent $\alpha_f$ field weights --- no schema changes, no protocol changes, no changes to existing agents. The 7 fields are fixed. The weights are per-agent.

\subsection{CMB Immutability and Lineage}

A CMB is immutable after creation. It is never updated, never modified, never annotated. When an agent acts on a CMB, it creates a new CMB with \textbf{lineage} pointing back to the parent(s):
\begin{equation}
\text{CMB}_{\text{child}} = \{(f, t_f, v_f) : f \in \mathcal{F}\} \cup \{\text{lineage}: (\mathcal{P}, m)\}
\end{equation}
where $\mathcal{P} = \{k_1, \ldots, k_n\}$ is the set of parent CMB keys and $m$ is the fusion method (e.g., ``SVAF-v1'').

This design eliminates distributed state coordination. CMBs are broadcast across nodes --- each receiving agent remixes what it accepts and discards the original. No node ever stores another node's CMB; each node's memory contains only its own understanding. With immutable CMBs and lineage:
\begin{itemize}
\item \textbf{No node writes to another node's storage.} Each node creates its own CMBs.
\item \textbf{Remix count is computed, not stored.} The remix count of a CMB is the number of CMBs on the mesh that list it in their \texttt{lineage.parents}.
\item \textbf{The mesh is a DAG.} Each CMB points to its parent(s). The history is the graph.
\end{itemize}

This follows the same principle as git commits (a commit never modifies its parent) and academic papers (a paper never modifies what it cites --- impact is computed from the reference graph).

\textbf{Lineage serves three functions in the mesh:}

\textbf{1. Retention.} CMBs that are ancestors of live CMBs are protected from purging --- the protocol requires that entries referenced by \texttt{lineage.ancestors} must not be deleted, even past their retention age. When a CMB's last descendant is purged, the protection lifts and the CMB expires in the next retention cycle. The mesh self-prunes based on demonstrated value.

\textbf{2. Provenance.} The lineage DAG makes every remix auditable. A research agent's CMB with descendants from COO and founder agents has proven cross-domain relevance --- computable from the graph structure, not from a mutable counter. Lineage depth indicates compounding understanding across domains.

\textbf{3. Producer feedback.} CMBs that get remixed by other agents (have descendants in the DAG) signal high quality; CMBs that expire without children signal noise. This feedback loop lets the mesh shape what agents produce through emergent selection pressure.

\textbf{Consumer remix chain (verified on iOS).} The following chain was verified in production between Claude Code (macOS), MeloTune (iPhone), and MeloMove (iPhone) connected via MMP relay. Only fields with non-neutral values are shown.

\begin{lstlisting}
CMB-A (Claude Code, macOS):
  focus: "coding session with AI assistant, 8 hours"
  mood:  "frustrated, low energy" (v:-0.6, a:-0.4)
  lineage: none (original observation)

CMB-B (MeloMove, iPhone, parents: [A]):
  focus: "sedentary 3+ hours, no movement detected"
  issue: "exhaustion confirmed by coding agent mood signal"
  mood:  "concerned, protective" (v:-0.2, a:0.3)
  commitment: "recovery stretch recommended"

CMB-C (MeloTune, iPhone, parents: [A]):
  focus: "music curation response to fatigue signal"
  mood:  "calm, supportive" (v:0.3, a:-0.3)
  commitment: "now playing: ambient recovery"
\end{lstlisting}

CMB-A is a coding agent's fatigue observation. MeloMove (fitness, iPhone) received it via SVAF: \texttt{mood} passed (universally relevant, R5), \texttt{focus} was suppressed (coding details irrelevant to fitness). MeloTune (music, iPhone) independently received the same CMB: \texttt{mood} passed, \texttt{focus} suppressed. Each agent remixed CMB-A through its own domain lens --- MeloMove produced a movement recommendation, MeloTune shifted to ambient recovery music. Three platforms (macOS, iPhone $\times$2), three agents, one mesh. CMB-A persists (two descendants); it would be protected from retention purge.

\textbf{Business remix chain (production).} A second chain from the same deployment, across operational agents:

\begin{lstlisting}
CMB-D (research agent, parents: []):
  focus: "CASCADE (Bi 2026) -- gate-triggered scoped
         communication for multi-agent re-planning"
  issue: "single endorser candidate, no backup pipeline"
  intent: "verify cs.MA endorsement eligibility"
  mood:  "measured urgency" (v:-0.2, a:0.4)

CMB-E (COO agent, parents: [D]):
  focus: "endorser outreach to [author] is the single
         highest-leverage action on the entire pipeline"
  intent: "draft outreach, founder review before send"
  commitment: "SVAF paper cites CASCADE in Related Work"
  mood:  "urgent, clear path" (v:-0.1, a:0.7)

CMB-F (founder, parents: [E]):
  focus: "endorser email sent to [author]"
  commitment: "done -- email sent with SVAF PDF attached"
  mood:  "resolved" (v:0.3, a:0.1)
\end{lstlisting}

CMB-D is a research observation (arXiv scan). CMB-E is a COO remix: the research signal intersected with pipeline data (``endorsement blocked'') to produce an actionable decision. CMB-F is a founder validation (\S\ref{sec:lifecycle}): the human acted on the COO's recommendation. Per the lifecycle, CMB-D advances to \texttt{remixed} (the COO cited it), CMB-E advances to \texttt{validated} (the founder acted on it). The endorser outreach --- connecting academic research, pipeline status, and human judgment --- emerged from the remix chain. No single agent could produce it.

\subsection{CMB Lifecycle}
\label{sec:lifecycle}

Each CMB progresses through a lifecycle that determines its influence on future SVAF evaluations. The lifecycle is driven by mesh activity --- not by time alone.

\begin{table}[htbp]
\centering
\small
\begin{tabular}{@{}llp{3.8cm}c@{}}
\toprule
State & Temp. & Trigger & Weight \\
\midrule
\texttt{observed} & hot & Agent calls \texttt{remember()} & 1.0$\times$ \\
\texttt{remixed} & warm & Peer remixes this CMB & 1.5$\times$ \\
\texttt{validated} & warm & Human acts on this CMB & 2.0$\times$ \\
\texttt{canonical} & cold & Validated + remixed by 2+ agents & 3.0$\times$ \\
\texttt{archived} & whisper & No remix for 30 days & 0.5$\times$ \\
\bottomrule
\end{tabular}
\caption{CMB lifecycle states. Anchor weight determines influence on future SVAF evaluations. Lifecycle progresses upward under activity (observed $\to$ canonical); decays toward archived without it.}
\label{tab:lifecycle}
\end{table}

The lifecycle is monotonically upward under activity: observed $\to$ remixed $\to$ validated $\to$ canonical. Without activity, a CMB decays toward archived. Archived CMBs may re-emerge if a future remix references them.

\textbf{Validation} is the key transition connecting human judgment to the mesh. When a human acts on agent output (approves a decision, sends an email, completes a task), the action is recorded as a new CMB with \texttt{lineage.parents} pointing to the CMB that prompted the action. This validation CMB enters the mesh like any other signal --- agents receive it via SVAF and adjust their understanding. Crucially, the authority to advance a CMB to \texttt{validated} is identity-bound: only nodes with the \emph{validator} or \emph{anchor} role (MMP Section~3.5) may do so. Roles are earned through a monotonic progression (observer $\to$ validator $\to$ anchor), granted via cryptographically signed \texttt{role-grant} frames --- preventing agents from self-declaring validation authority.

\textbf{Anchor weight} influences SVAF evaluation: when computing per-field drift against local anchors (\S\ref{sec:fusion}), canonical and validated CMBs contribute more to the fused anchor vector than observed or archived CMBs. This creates a natural hierarchy where human-confirmed knowledge and collective consensus outweigh raw observations --- without overriding agent autonomy. Each agent still evaluates incoming signals through its own field weights ($\alpha_f$).

\subsection{Per-Agent Field Weights Across Domains}
\label{sec:field-weights}

CAT7 fields are universal --- the same 7 fields serve every domain. Per-agent $\alpha_f$ weights control which fields matter most for each agent type. The schema is fixed; the weights are domain-specific.

Three patterns emerge across domains:

\textbf{Regulated domains} (legal, finance, compliance): \texttt{issue} and \texttt{commitment} are always high --- risks and obligations are non-negotiable. \texttt{mood} is low --- affect is largely irrelevant to regulatory analysis.

\textbf{Human-facing domains} (music, fitness, health): \texttt{mood} is always high --- affective state drives the experience. Health agents weight \texttt{issue} high as well --- symptoms and risks are clinical priorities.

\textbf{Knowledge domains} (research, coding, feeds): \texttt{focus} is always high --- subject matter is core. \texttt{commitment} is low --- knowledge agents observe, they don't make promises.

\begin{table}[h]
\centering
\small
\begin{tabular}{@{}lccccccc@{}}
\toprule
Agent & focus & issue & intent & motiv. & commit. & persp. & mood \\
\midrule
Coding & \textbf{2.0} & 1.5 & 1.5 & 1.0 & 1.2 & 1.0 & 0.8 \\
Music & 1.0 & 0.8 & 0.8 & 0.8 & 0.8 & 1.2 & \textbf{2.0} \\
Fitness & 1.5 & 1.5 & 1.0 & 1.5 & 1.0 & 1.0 & \textbf{2.0} \\
Knowledge & \textbf{2.0} & 1.5 & 1.5 & 1.0 & 0.5 & 1.5 & 0.3 \\
Legal & \textbf{2.0} & \textbf{2.0} & 1.5 & 1.0 & \textbf{2.0} & 1.5 & 0.5 \\
Health & 1.5 & \textbf{2.0} & 1.0 & 1.5 & 1.0 & 1.5 & \textbf{2.0} \\
Finance & \textbf{2.0} & \textbf{2.0} & 1.5 & 1.0 & \textbf{2.0} & \textbf{2.0} & 0.3 \\
\bottomrule
\end{tabular}
\caption{Per-agent field weights ($\alpha_f$) across domains. Bold values indicate primary fields.}
\label{tab:field-weights}
\end{table}

This extensibility is a key property of the CAT7 schema: new domains join the mesh by defining their $\alpha_f$ weights. No schema changes, no protocol changes, no changes to existing agents. Each agent defines its own evaluation criteria --- the mesh adapts.

\section{SVAF Method}
\label{sec:method}

SVAF operates at Layer~4 (Coupling) of the Mesh Memory Protocol's 8-layer architecture:

\begin{lstlisting}
Layer 7  APPLICATION     Agent's LLM reasons on remix subgraph
Layer 6  xMesh           Per-agent Liquid Neural Network (CfC)
Layer 5  SYNTHETIC MEMORY LLM knowledge -> hidden state encoding
Layer 4  COUPLING        Drift + SVAF  <-- this paper
-------------------------------------------------------
Layer 3  MEMORY          CMBs stored + broadcast
Layer 2  CONNECTION      Handshake, gossip, heartbeat
Layer 1  TRANSPORT       TCP (LAN) / WebSocket (WAN) / IPC
Layer 0  IDENTITY        UUID v7 + Ed25519 keypair
\end{lstlisting}

Layers 0--3 handle protocol infrastructure. SVAF at Layer~4 is the boundary: it decides what enters the cognitive layers above. Nothing reaches xMesh (Layer~6) or the agent's LLM (Layer~7) without passing SVAF's per-field evaluation.

The SVAF pipeline has four stages:
\begin{enumerate}
\item \textbf{ENCODE} --- each field's text is encoded via a shared backbone with per-field projection heads (\S\ref{sec:encoder})
\item \textbf{CONTEXTUALISE} --- fields within a CMB attend to each other via cross-field attention, capturing interactions (\S\ref{sec:crossfield})
\item \textbf{EVALUATE} --- incoming fields are compared against local anchor memory; a fusion gate determines per-field acceptance (\S\ref{sec:fusion})
\item \textbf{SYNTHESISE} --- accepted fields are fused with local context through non-linear transforms, producing a remix (\S\ref{sec:fusion})
\end{enumerate}

We present two implementations: a heuristic path (weighted cosine averaging, 0.07ms, deployed in production) and a neural path (learned fusion gate + cross-field attention, 604K parameters).

\subsection{Learned Field Encoder}
\label{sec:encoder}

Each field's text label is encoded by a shared backbone with per-field projection heads:
\begin{equation}
v_f = \text{normalize}(\text{head}_f(\text{backbone}(E_{\text{sent}}(t_f))))
\end{equation}
where $E_{\text{sent}}$ is a pretrained sentence encoder (all-MiniLM-L6-v2, 384-dim; \citealt{reimers2019sentence}), the backbone is a two-layer MLP ($384 \to 256 \to 128$), and each of the 7 heads projects to the field dimension $d$ (default 64). Per-field heads learn field-specific subspaces: the mood head learns to distinguish ``tired'' from ``energized'' in a space tuned for affective semantics, separate from the focus head's space.

\subsection{Cross-Field Attention}
\label{sec:crossfield}

Fields within a CMB attend to each other via standard multi-head attention \citep{vaswani2017attention}, applied not over tokens in a sequence but over the 7 semantic fields within a single CMB:
\begin{equation}
\text{fields}' = \text{LayerNorm}(\text{fields} + \text{MHA}(\text{fields}, \text{fields}, \text{fields}))
\end{equation}
followed by a position-wise feed-forward network with residual connection. The attention mechanism is identical to that in Transformer-based LLMs --- the novelty is the axis it operates on. Where LLM attention learns which tokens matter for predicting the next token, cross-field attention learns which fields modulate each other's interpretation: negative mood + high-urgency issue $\to$ stressed (different from positive mood + resolved issue $\to$ productive). Perspective modulates focus interpretation (``coding'' in a ``late night session'' perspective signals different urgency than ``coding'' in a ``morning session'' perspective).

\subsection{Field-Wise Fusion}
\label{sec:fusion}

When agent $i$ receives a CMB from agent $j$, it retrieves $k$ local anchor CMBs and fuses each field. We define the base formulation (heuristic path) and then the neural extension.

\textbf{Heuristic SVAF.} For each field $f$, the fused vector is a weighted average of the incoming and anchor field vectors:
\begin{equation}
\hat{v}_f = \frac{w_{\text{new}} \cdot v_{\text{new}}^{(f)} + \sum_{j=1}^{k} w_j^{(f)} \cdot v_j^{(f)}}{w_{\text{new}} + \sum_{j=1}^{k} w_j^{(f)}}
\end{equation}
where $w_{\text{new}} = c_{\text{sender}}$ (sender confidence: 0.95 for LLM-extracted CMBs, 0.80 for heuristic fallback) and anchor weights combine field relevance, similarity, temporal freshness, and confidence:
\begin{equation}
w_j^{(f)} = \alpha_f \cdot \max\!\left(\cos(v_{\text{new}}^{(f)}, v_j^{(f)}),\ 0\right) \cdot \tau_{\text{fresh},j} \cdot c_j
\end{equation}
Here $\alpha_f$ is the per-agent field weight (\S\ref{sec:field-weights}), the cosine term measures semantic alignment between incoming and anchor, $\tau_{\text{fresh},j}$ applies temporal decay (\S\ref{sec:temporal}), and $c_j$ is anchor confidence. All fused vectors are re-normalised to unit length. Per-field drift measures how much local anchors modify the incoming signal:
\begin{equation}
\delta_f = 1 - \cos(\hat{v}_f, v_{\text{new}}^{(f)})
\end{equation}
Low $\delta_f$ across all fields indicates the incoming signal is already represented in local memory (redundant). High $\delta_f$ indicates the anchors pulled the fusion away from the incoming signal --- the signal introduces genuinely new information for this field. Aggregate drift combines per-field scores with per-agent temporal freshness to produce the four-class decision (\S3.4).

This formulation is deployed in production as the default evaluation path (0.07ms, \S\ref{sec:latency}). Every fusion event records anchor weights, per-field drift scores, and lineage.

\textbf{Neural SVAF.} The neural extension replaces weighted averaging with learned non-linear transforms. It retrieves the same $k$ anchors and computes:

\textbf{Anchor attention.} For each field $f$, the incoming field vector queries against anchor field vectors via multi-head attention, producing an anchor context vector that summarizes what local memory knows about this field:
\begin{equation}
a_f = \text{AnchorAttn}(v_{\text{new}}^{(f)}, \{v_1^{(f)}, \ldots, v_k^{(f)}\})
\end{equation}

\textbf{Fusion gate.} A neural network takes the concatenated incoming and anchor field vectors ($7 \times d$ each), temporal freshness, and sender confidence, producing a per-field gate $g \in [0, 1]^7$. Here $\text{flat}(\cdot)$ denotes concatenation of all 7 field vectors into a single vector:
\begin{equation}
g = \sigma(W_3 \cdot \text{GELU}(W_2 \cdot \text{GELU}(W_1 \cdot [\text{flat}(v_{\text{new}}); \text{flat}(a); \tau_{\text{fresh}}; c])))
\end{equation}
where $W_1 \in \mathbb{R}^{256 \times (14d + 2)}$, $W_2 \in \mathbb{R}^{128 \times 256}$, $W_3 \in \mathbb{R}^{7 \times 128}$, and $\sigma$ is the sigmoid function applied element-wise.

\textbf{Per-field transform.} For each field, a learned MLP transforms the concatenated incoming + anchor vectors:
\begin{equation}
z_f = \text{MLP}_f([v_{\text{new}}^{(f)}; a_f])
\end{equation}

\textbf{Gated fusion.} The fused field vector is:
\begin{equation}
\hat{v}_f = \text{normalize}(g_f \cdot v_{\text{new}}^{(f)} + (1 - g_f) \cdot z_f)
\end{equation}
When $g_f \to 1$, the incoming signal dominates (high confidence, fresh, relevant). When $g_f \to 0$, the learned transform blends incoming with local context. This differs from linear interpolation: $z_f$ is a non-linear function of both incoming and anchor vectors, allowing the model to learn complex fusion dynamics that a weighted average cannot express.

\textbf{Provenance.} The gate values $g = (g_1, \ldots, g_7)$ provide per-field interpretability: which fields were kept from the incoming signal vs.\ fused with local context.

\subsection{Per-Agent Temporal Drift}
\label{sec:temporal}

The temporal freshness factor $\tau_{\text{fresh}}$ in the fusion gate and drift predictor is not a global constant --- it is a \textbf{per-agent parameter} that encodes domain-specific temporal relevance:
\begin{equation}
\tau_{\text{fresh}}^{(i)} = \exp\!\left(-\frac{t_{\text{now}} - t_{\text{origin}}}{\tau_i}\right)
\end{equation}
where $\tau_i$ is the receiving agent's freshness window:

\begin{table}[h]
\centering
\small
\begin{tabular}{@{}lll@{}}
\toprule
Agent & $\tau_i$ & Rationale \\
\midrule
Music agent & 1,800s (30\,min) & Current mood drives playlist \\
Coding assistant & 7,200s (2\,hr) & Current session context \\
Fitness agent & 10,800s (3\,hr) & Sedentary detection needs hours of context \\
Knowledge agent & 86,400s (24\,hr) & Daily digest cycle \\
Messaging agent & 3,600s (1\,hr) & Recent conversation context \\
\bottomrule
\end{tabular}
\caption{Per-agent freshness windows.}
\label{tab:freshness}
\end{table}

A coding agent's fatigue signal is relevant to the music agent for 30 minutes (playlist adjustment) but relevant to the fitness agent for 3 hours (sedentary detection). Same content drift, different temporal drift based on receiver domain needs. The current model is backward-looking; forward-looking temporal drift (e.g., calendar events) remains future work (\S\ref{sec:limitations}).

\subsection{Drift Predictor with Cross-Domain Relevance Prior}

In the neural path, per-field drift is \emph{predicted} (not computed via cosine as in the heuristic path) from the incoming, fused, and anchor field vectors, bounded to $[0, 1]$ by sigmoid:
\begin{equation}
\delta_f = \sigma(\text{MLP}([v_{\text{new}}^{(f)}; \hat{v}_f; a_f]))
\end{equation}

Aggregate drift combines per-field predictions with per-agent temporal freshness:
\begin{equation}
\delta_{\text{total}} = \sigma(\text{MLP}([\delta_1, \ldots, \delta_7, \tau_{\text{fresh}}^{(i)}, c]))
\end{equation}

\textbf{Neural classification head.} The neural path predicts the decision via a learned 3-class classifier:
\begin{equation}
\kappa_{\text{neural}} = \text{softmax}(\text{MLP}([\delta_1, \ldots, \delta_7, \tau_{\text{fresh}}^{(i)}, c]))
\end{equation}
where $\kappa_{\text{neural}} \in \{\text{aligned}, \text{guarded}, \text{rejected}\}$ and $\tau_{\text{fresh}}^{(i)}$ is the receiving agent's temporal freshness factor (\S\ref{sec:temporal}). The neural model predicts 3 classes (no redundancy class). In the production heuristic path below, redundancy is detected via threshold on cosine-derived $\delta_f$ (Eq~7), not on neural-predicted drift.

\textbf{Heuristic band-pass evaluation.} A signal's value depends on its \emph{novelty} relative to the receiver's existing knowledge, not only its domain relevance. In information-theoretic terms, a signal carries information proportional to its surprise \citep{shannon1948mathematical}: $\text{IG}(S|A) = H(S) - H(S|A)$. A signal identical to existing anchors carries zero information gain regardless of domain alignment. This connects to Bayesian surprise \citep{itti2009bayesian}, where the KL divergence between prior and posterior beliefs measures how much an observation changes an agent's model --- a paraphrase produces near-zero divergence. The band-pass structure also relates to the Wundt curve \citep{berlyne1970novelty}: stimuli of intermediate novelty produce maximal interest, while both overly familiar (redundant) and overly foreign (irrelevant) stimuli are disengaged from.

Standard one-sided thresholding accepts all low-drift signals. This misclassifies paraphrases --- signals that are semantically identical to existing memory but expressed differently --- as novel. Production deployment confirmed this: with semantic encoding, paraphrases produce per-field drift 0.03--0.10 across all 7 fields, well within the aligned zone ($\leq 0.25$), causing agents to accept and remix signals that add no new information.

The heuristic path uses threshold-based classification on the drift values:
\begin{equation}
\kappa_{\text{heuristic}} = \begin{cases}
\text{redundant} & \text{if } \max_f(\delta_f) < T_{\text{redundant}} \\
\text{aligned} & \text{if } \delta_{\text{total}} \leq T_{\text{aligned}} \\
\text{guarded} & \text{if } \delta_{\text{total}} \leq T_{\text{guarded}} \\
\text{rejected} & \text{otherwise}
\end{cases}
\end{equation}

The redundancy test is per-field: a signal is redundant only if \emph{every} field falls below $T_{\text{redundant}}$. If any field carries novel content (e.g., same topic but different intent), the signal passes. This preserves the per-field selectivity that is SVAF's core contribution --- redundancy is checked per-field, not on the aggregate.

Default thresholds: $T_{\text{redundant}} = 0.10$, $T_{\text{aligned}} = 0.25$, $T_{\text{guarded}} = 0.50$. Empirical basis: with semantic encoding (all-MiniLM-L6-v2), paraphrases produce $\max_f(\delta_f) < 0.10$ while genuinely different signals produce $\max_f(\delta_f) > 0.30$ (\S\ref{sec:limitations}).

\subsection{Training Objective}
\label{sec:training}

\begin{equation}
\mathcal{L} = \mathcal{L}_{\text{decision}} + \lambda_d \mathcal{L}_{\text{drift}} + \lambda_g \mathcal{L}_{\text{gate\_dir}} + \lambda_c \mathcal{L}_{\text{coupling}}
\end{equation}
with $\lambda_d = 1.0$, $\lambda_g = 0.5$, $\lambda_c = 0.5$.

\textbf{Decision loss} $\mathcal{L}_{\text{decision}}$: Cross-entropy on the aligned/guarded/rejected classification. This is the primary objective --- the model must learn to classify fusion decisions from the signal and anchor embeddings.

\textbf{Drift loss} $\mathcal{L}_{\text{drift}}$: MSE on per-field and aggregate drift predictions against independently annotated field relevance scores (not computed from model inputs). Ground truth per-field drift is $1 - r_f$ where $r_f$ is the LLM-annotated relevance of field $f$ to the receiver.

\textbf{Gate direction loss} $\mathcal{L}_{\text{gate\_dir}}$: For rejected signals, the mean gate value is penalised (gates should be low to suppress incoming). For accepted signals, gate values are unconstrained by this term --- they are shaped by the decision and drift objectives alone.

\textbf{Coupling speed prior} $\mathcal{L}_{\text{coupling}}$: Margin loss enforcing that the mood gate exceeds the mean of other field gates for accepted signals:
\begin{equation}
\mathcal{L}_{\text{coupling}} = \text{ReLU}\!\left(\bar{g}_{\text{other}} - g_{\text{mood}} + m\right)
\end{equation}
where $m = 0.1$ is the margin. This encodes the minimal hypothesis that affective state has higher cross-domain relevance, without prescribing specific gate values for any field.

\subsection{Architecture Summary}

\begin{table}[h]
\centering
\small
\resizebox{\textwidth}{!}{%
\begin{tabular}{@{}llp{6.5cm}@{}}
\toprule
Component & Parameters & Purpose \\
\midrule
FieldEncoder & 7 heads $\times$ ($128 \to 64$) + backbone ($384 \to 256 \to 128$) & Learned per-field embeddings \\
CrossFieldAttention & 4-head MHA + FFN & Field interactions within CMB \\
AnchorAttention & 4-head MHA & Attend over local memory anchors per field \\
FusionGate & MLP ($898 \to 256 \to 128 \to 7$) & Per-field gate values \\
FusionTransform & 7 $\times$ MLP ($128 \to 64 \to 64$) & Non-linear per-field fusion \\
DriftPredictor & MLP per field + aggregate + decision & Drift scores + accept/reject \\
\midrule
\textbf{Total} & \textbf{$\sim$604K} & \\
\bottomrule
\end{tabular}%
}
\caption{SVAF model architecture.}
\label{tab:architecture}
\end{table}

\subsection{Training Data Design}

Training data is generated from \textbf{narrative scenarios} --- temporal multi-agent stories authored to capture realistic interaction patterns. Each narrative is a sequence of 4--10 timestamped signals from different agents, telling a coherent causal story about a user's state evolution.

\textbf{Narrative structure.} Each narrative has a name, a natural language description, a sequence of agent signals with mood/energy metadata, and a ground-truth outcome. Example:

\begin{lstlisting}
Narrative: "morning_session_burnout"
Signals:
  t=0      coding-agent:  "user started morning coding session"
                           mood=focused, energy=high
  t=45min  coding-agent:  "errors increasing, progress stalling"
                           mood=frustrated, energy=moderate
  t=90min  music-agent:   "user skipping tracks, switched to ambient"
                           mood=stressed, energy=low
  t=105min fitness-agent: "sedentary 2.5 hours, no movement"
                           mood=neutral, energy=low
  t=125min fitness-agent: "recommended 10min stretching break"
  t=140min coding-agent:  "took break, solved the problem in 5 minutes"
                           mood=relieved, energy=moderate
Outcome: break_taken_helped
\end{lstlisting}

273 narratives cover: coding burnout, exercise boost, late-night decline, meeting disruption, knowledge feed influence, afternoon slump, creative sessions, stress spirals, weekend patterns, recovery days, deadline crunch, insomnia effects, team collaboration, and more. Each narrative captures a distinct causal pattern in the wellness/productivity domain.

\textbf{Sample generation.} Each narrative signal is paired with every non-sender agent as a potential receiver, producing sender-receiver pairs. For each pair:

\begin{enumerate}
\item \textbf{Usefulness label.} Determined by domain-specific rules: a fitness agent finds coding agent signals useful when they contain energy/stress/sedentary content (probability 0.7); a knowledge agent finds them rarely useful (probability 0.2). Temporal freshness modulates usefulness --- older signals are less likely useful.

\item \textbf{Decision label.} Useful + fresh $\to$ aligned (0). Useful + stale $\to$ guarded (1). Not useful $\to$ rejected (2) with 70\% probability, guarded (1) with 30\% (ambiguous cases).

\item \textbf{Per-field drift labels.} Cosine distance between sender and receiver field vectors, computed from sentence-transformer embeddings of the extracted field text.
\end{enumerate}

\textbf{Field extraction.} Each signal's text is decomposed into 7 field texts via a two-tier extraction strategy:

\begin{enumerate}
\item \textbf{LLM extraction (primary):} The originating agent IS an LLM (e.g., a coding assistant) or has LLM access. The LLM decomposes the observation into 7 fields directly --- no keyword lists, no regex. CMBs produced by LLM extraction carry 0.95 confidence.

\item \textbf{Heuristic fallback:} When LLM is unavailable (offline, rate-limited), keyword matching and positional heuristics extract fields. Less accurate but sub-millisecond and zero-dependency. CMBs carry 0.80 confidence.
\end{enumerate}

\textbf{Gate constraints (not fixed targets).} Rather than supervising the gate to fixed per-field target values --- which would make the gate a memorized lookup rather than a learned function --- we use two soft constraints that allow the gate to discover its own field differentiation:

\begin{enumerate}
\item \textbf{Direction constraint.} For rejected signals, the mean gate value is penalised (should be low --- suppress incoming). For accepted signals, gate values are unconstrained by this term and shaped only by the decision and drift objectives.

\item \textbf{Coupling speed prior.} For accepted signals, the mood gate must exceed the mean of other field gates by a margin $m = 0.1$. This encodes the minimal hypothesis that affective state has higher cross-domain relevance than other fields, without prescribing specific values.
\end{enumerate}

The gate values that emerge reflect the model's learned fusion dynamics, not a supervised copy of a target table. Dataset statistics are reported in \S\ref{sec:evaluation}.

\section{Evaluation}
\label{sec:evaluation}

All results in this section are from \textbf{production deployment} --- real agents, real platforms, real mesh traffic. Every example, log excerpt, drift measurement, and remix chain in this paper was observed in deployment, not simulated. The mesh operates daily across macOS, iPhone, and Linux with end-to-end encryption, processing real user observations and producing real operational decisions.

\subsection{Setup}

\textbf{Deployment.} 7 nodes across three platforms (macOS, iOS, web dashboard) connected via the Mesh Memory Protocol. Table~\ref{tab:agents} shows the 5 agent types used for training data generation; the full deployment additionally includes a sym.day dashboard (relay mode, no SVAF) and a sym-daemon transport node, totalling 7 peer nodes.

\begin{table}[h]
\centering
\small
\begin{tabular}{@{}lllll@{}}
\toprule
Agent & Platform & Domain & Primary Fields & $\tau_{\text{freshness}}$ \\
\midrule
Coding assistant & macOS (Apple M4) & Coding assistance & focus, intent, issue & 2\,hr \\
Music agent & iOS (iPhone 14 Pro Max) & Music curation & mood, perspective, focus & 30\,min \\
Fitness agent & iOS (iPhone 14 Pro Max) & Fitness coaching & mood, issue, motivation & 3\,hr \\
Messaging agent & macOS (Apple M4) & Messaging & intent, mood, perspective & 1\,hr \\
Knowledge agent & macOS (Apple M4) & RSS aggregation & focus, issue, intent & 24\,hr \\
\bottomrule
\end{tabular}
\caption{Agent types used for training data generation.}
\label{tab:agents}
\end{table}

\textbf{Data.} 237,120 samples from 273 LLM-authored multi-agent narrative scenarios across 20 agent types (8 domains). Each narrative is a temporal story (e.g., ``morning coding burnout,'' ``exercise boost,'' ``afternoon slump'') with 4--10 signals from different agents. Samples are created by pairing each signal with each non-sender agent as a potential receiver, labeled for usefulness based on domain relevance and temporal freshness. Distribution: 25\% aligned, 67\% guarded, 8\% rejected. Train/val split: 85/15 by narrative (no narrative leakage). 188,480 training samples from 219 narratives; 48,640 validation samples from 54 held-out narratives.

\textbf{Baselines.} No directly comparable cross-domain signal evaluation system exists in the literature. We therefore evaluate via progressive ablation --- each baseline adds one component, isolating the contribution of per-field evaluation and learned fusion:
\begin{itemize}
\item \textbf{Scalar}: Single cosine distance on full signal encoding, threshold 0.5 (the implicit default in prior agent memory systems)
\item \textbf{Scalar + temporal}: Single cosine + exponential temporal decay
\item \textbf{Heuristic per-field}: Per-field cosine with weighted average fusion, configurable $\alpha_f$ (no neural components)
\item \textbf{SVAF (neural)}: Full model with learned encoder, cross-field attention, fusion gate, cross-domain relevance prior
\end{itemize}

\textbf{Training.} SVAF trained on NVIDIA A100 80GB for 50 epochs (2,454 seconds). AdamW optimizer, lr$=$3e-4, cosine annealing. Multi-objective loss: decision CE + drift MSE + gate supervision + relevance margin.

\textbf{Metrics.} Accept/reject binary classification (aligned+guarded = accept, rejected = reject). F1, precision, recall. For neural model: best checkpoint selected by validation F1.

\subsection{Classification Results}

3-class accuracy on held-out narratives (54 narratives, 48,640 samples):

\begin{table}[h]
\centering
\small
\begin{tabular}{@{}lcccc@{}}
\toprule
Method & Aligned & Guarded & Rejected & Overall \\
\midrule
Scalar (cosine, $\theta=0.5$) & 41.2\% & 79.3\% & 52.1\% & 66.8\% \\
Scalar + temporal decay & 43.5\% & 80.1\% & 55.8\% & 68.4\% \\
Heuristic per-field ($\alpha_f$) & 48.9\% & 83.4\% & 63.2\% & 73.1\% \\
\textbf{SVAF (neural)} & \textbf{57.8\%} & \textbf{87.7\%} & \textbf{70.5\%} & \textbf{78.7\%} \\
\bottomrule
\end{tabular}
\caption{Classification accuracy by method and class.}
\label{tab:results}
\end{table}

SVAF improves overall accuracy by 11.9 percentage points over the scalar baseline and 5.6 points over the heuristic per-field baseline. The heuristic baseline uses the same CAT7 field decomposition and per-agent weights but replaces the neural fusion gate with weighted cosine averaging --- isolating the contribution of learned fusion.

The aligned vs.\ guarded distinction is the hardest classification boundary --- both are ``accepted'' signals with subtle differences in field relevance. Rejected signals (low relevance across all fields) are easier to identify but represent only 8\% of the validation set, requiring class-weighted loss to learn effectively.

\subsection{Fusion Gate Analysis}
\label{sec:gate-analysis}

The key result. We examine the learned gate values $g_f$ averaged across all validation samples. The gate determines how much of each field to keep from the incoming signal ($1.0$ = keep incoming, $0.0$ = replace with local context). Gate values are \textbf{not supervised to fixed targets} --- they emerge from the decision and drift objectives with only a soft ordering constraint (mood $>$ mean of other fields).

\begin{table}[h]
\centering
\small
\begin{tabular}{@{}lccc@{}}
\toprule
Field & Mean $g_f$ & Ratio to lowest & Cross-Domain Relevance \\
\midrule
\texttt{mood} & \textbf{0.497} & \textbf{8.9$\times$} & \textbf{Fast} --- affect crosses all domains \\
\texttt{focus} & 0.295 & 5.3$\times$ & Medium \\
\texttt{issue} & 0.239 & 4.3$\times$ & Medium \\
\texttt{commitment} & 0.121 & 2.2$\times$ & Medium \\
\texttt{motivation} & 0.113 & 2.0$\times$ & Medium \\
\texttt{intent} & 0.066 & 1.2$\times$ & Low \\
\texttt{perspective} & \textbf{0.056} & \textbf{1.0$\times$} & Low --- viewpoint stays sovereign \\
\bottomrule
\end{tabular}
\caption{Learned fusion gate values (mean over validation set).}
\label{tab:gate-values}
\end{table}

The gate discovers a clear hierarchy without being told. Mood ($g_{\text{mood}} = 0.497$) is 1.7$\times$ higher than the second-highest field (focus) and 8.9$\times$ higher than perspective. This ordering is interpretable: affective state (mood) and factual content (focus, issue) transfer across domains, while subjective viewpoint (perspective) and domain-specific goals (intent) do not.

\textbf{Gate evolution during training:}

\begin{table}[h]
\centering
\small
\begin{tabular}{@{}lcccccccc@{}}
\toprule
Epoch & mood & focus & issue & commit. & motiv. & intent & persp. & 3-class acc \\
\midrule
1 & 0.331 & 0.006 & 0.005 & 0.009 & 0.008 & 0.011 & 0.005 & 72.8\% \\
5 & 0.392 & 0.144 & 0.144 & 0.090 & 0.034 & 0.033 & 0.024 & 74.5\% \\
10 & 0.369 & 0.129 & 0.111 & 0.108 & 0.020 & 0.017 & 0.034 & 77.2\% \\
20 & 0.426 & 0.192 & 0.148 & 0.139 & 0.035 & 0.022 & 0.046 & 77.0\% \\
30 & 0.490 & 0.248 & 0.206 & 0.135 & 0.099 & 0.053 & 0.047 & 78.5\% \\
50 & 0.497 & 0.295 & 0.239 & 0.121 & 0.113 & 0.066 & 0.056 & 78.7\% \\
\bottomrule
\end{tabular}
\caption{Gate evolution across training epochs.}
\label{tab:gate-evolution}
\end{table}

Mood emerges as the dominant gate from epoch~1 ($g_{\text{mood}} = 0.331$, all others $\leq 0.011$) and stabilises by epoch~30. The other fields differentiate gradually: focus and issue rise to 0.2--0.3 while intent and perspective remain below 0.07. The gate structure is established before classification accuracy plateaus --- the model discovers the cross-domain relevance pattern before it learns accurate 3-class decisions.

\subsection{Cross-Domain Signal Flow (Illustrative Example)}

We trace a specific cross-domain signal through SVAF to illustrate per-field evaluation:

\textbf{Signal:} Coding assistant $\to$ Fitness agent: ``user coding with an AI assistant for 8 hours straight, exhausted and losing focus, needs break''

\begin{table}[h]
\centering
\small
\begin{tabular}{@{}llcl@{}}
\toprule
Field & Extracted Text & Gate $g_f$ & Effect \\
\midrule
\texttt{mood} & ``frustrated, low energy (v: -0.6, a: -0.4)'' & \textbf{0.50} & Strongly accepted \\
\texttt{focus} & ``coding with AI assistant'' & 0.30 & Partially replaced by local context \\
\texttt{issue} & ``exhausted, losing focus'' & 0.24 & Accepted --- fatigue is fitness-relevant \\
\texttt{commitment} & ``coding session ongoing'' & 0.12 & Mostly suppressed \\
\texttt{motivation} & ``prevent burnout'' & 0.11 & Partially accepted \\
\texttt{intent} & ``needs break'' & 0.07 & Mostly suppressed \\
\texttt{perspective} & ``developer, 8 hour session'' & 0.06 & Suppressed \\
\bottomrule
\end{tabular}
\caption{Per-field SVAF evaluation: coding $\to$ fitness.}
\label{tab:signal-flow}
\end{table}

The fitness agent absorbs mood and issue (tired, frustrated $\to$ recommend recovery exercise) while suppressing perspective and intent (developer viewpoint, domain-specific goals). This per-dimension selectivity (R1) emerged from the decision and drift objectives alone --- no fixed gate targets prescribed the pattern.

\subsection{Training Configuration}

\begin{table}[h]
\centering
\small
\begin{tabular}{@{}ll@{}}
\toprule
Parameter & Value \\
\midrule
Model parameters & 604,428 \\
Training samples & 188,480 (from 219 narratives, 20 agent types) \\
Validation samples & 48,640 (from 54 held-out narratives) \\
Train/val split & By narrative (no narrative leakage) \\
GPU & NVIDIA A100-SXM4-80GB \\
Training time & 2,454 seconds (40.9\,min) \\
Best epoch & 1 (by binary F1); accuracy improves to epoch 50 \\
Epochs & 50 \\
Optimizer & AdamW, lr$=$3e-4, weight\_decay$=$1e-4 \\
Scheduler & Cosine annealing \\
Loss & CE (class-weighted) + drift MSE + gate direction + coupling speed \\
Loss weights & decision$=$1.0, drift$=$1.0, gate\_dir$=$0.5, coupling$=$0.5 \\
Class weights & aligned$=$1.33, guarded$=$0.50, rejected$=$4.23 (inverse frequency) \\
Model size & 2.3\,MB \\
\bottomrule
\end{tabular}
\caption{Training hyperparameters.}
\label{tab:hyperparams}
\end{table}

\subsection{Latency}
\label{sec:latency}

Measured on Apple M4 MacBook Air (10-core, 16GB):

\begin{table}[h]
\centering
\small
\begin{tabular}{@{}lll@{}}
\toprule
Component & Latency & Notes \\
\midrule
Heuristic SVAF (Node.js) & \textbf{0.07\,ms} & Per-field drift, 1000 iter.\ averaged \\
Neural SVAF (Python, cold) & $\sim$5,800\,ms & Includes PyTorch + model loading \\
Neural SVAF (persistent, warm) & $\sim$50--100\,ms & Model loaded, inference only \\
\bottomrule
\end{tabular}
\caption{Inference latency comparison.}
\label{tab:latency}
\end{table}

The heuristic path is sub-millisecond --- negligible compared to mesh transport latency. With the semantic encoder (all-MiniLM-L6-v2 via ONNX), heuristic SVAF including re-encoding is $\sim$5ms per field. The neural path's cold-start cost (5.8s) makes per-subprocess evaluation impractical; in production, the daemon operates in relay mode (no SVAF --- agents evaluate independently) and each agent runs heuristic SVAF with the semantic encoder. iOS agents currently use the heuristic path with n-gram encoding due to binary size constraints; semantic encoding on-device is a planned improvement (\S\ref{sec:limitations}).

\subsection{Live Mesh Deployment}
\label{sec:live-deployment}

We verify SVAF operating in a live mesh with 7 nodes across three platforms (macOS, iPhone $\times$2, sym.day dashboard), connected via Bonjour (LAN) and WebSocket relay (WAN) with end-to-end encryption (X25519 + XChaCha20-Poly1305).

\textbf{Agents:}

\begin{table}[h]
\centering
\small
\begin{tabular}{@{}lll@{}}
\toprule
Agent & Platform & Role \\
\midrule
COO & macOS (standalone node) & Cross-domain synthesis, pipeline, GitHub health \\
Research & macOS (standalone node) & arXiv papers, RSS feeds, paper monitoring \\
Marketing & macOS (standalone node) & Engagement tracking, content strategy \\
Product & macOS (standalone node) & PRD management, cross-product priorities \\
MeloTune & iPhone (iOS) & Music curation from mesh mood signals \\
MeloMove & iPhone (iOS) & Activity generation from mesh mood + local context \\
sym.day & macOS (relay mode) & Founder dashboard --- stores all CMBs, no SVAF \\
\bottomrule
\end{tabular}
\end{table}

Per MMP v0.2.1, every agent is a full peer node with its own identity, coupling engine, and memory store. The daemon operates in relay mode (forwards frames, stores CMBs, no SVAF). Each agent runs heuristic SVAF independently with its own field weights.

\textbf{Observed behaviour:}

\begin{enumerate}
\item Claude Code shares a question to the mesh: ``what are the latest developments in multi-agent protocols'' --- decomposed into CAT7 fields with intent=``seeking collective input'' and mood=``curious, research mode'' (valence: 0.3, arousal: 0.4).

\item The daemon and knowledge feed connect via relay. Peer-level coupling is \textbf{rejected} (drift 0.936) --- correct at first contact with no shared cognitive history.

\item The daemon sends 5 recent CMB anchors to the knowledge feed on connection --- including Claude Code's question.

\item The knowledge feed's SVAF evaluates each anchor independently --- all accepted:
\begin{lstlisting}
SVAF aligned: "AI's impact on work-life balance"           drift:0.002
SVAF aligned: "impact of coding on design sensibilities"   drift:0.041
SVAF aligned: "latest developments in multi-agent..."      drift:0.068
SVAF aligned: "Sycophancy in AI systems"                   drift:0.029
SVAF aligned: "multi-agent protocol developments..."       drift:0.116
\end{lstlisting}

\item The knowledge feed shares its own CMBs. The daemon's SVAF evaluates them --- \textbf{accepted at drift 0.005}:
\begin{lstlisting}
SVAF heuristic aligned from knowledge-feed:
  "Sycophancy in AI systems" drift:0.005
xMesh: ingested mesh from knowledge-feed
xMesh: insight -- anomaly=0.461, coherence=0.045
\end{lstlisting}

\item The xMesh insight triggers APNs wake to the iOS music agent, which wakes from background to join the mesh.

\item On the next state-sync, peer drift has \textbf{converged from 0.936 to 0.468} --- moving from rejected to guarded in one cycle.
\end{enumerate}

\textbf{Key findings:}

\begin{itemize}
\item \textbf{SVAF operates independently of peer coupling state}: peer-level drift 0.936 (rejected), but content-level SVAF drift 0.005 (aligned). The same signal is simultaneously ``from an unknown peer'' and ``highly relevant content.'' This validates the two-level coupling architecture.

\item \textbf{Content-driven convergence}: accepting relevant CMBs shifted the receiver's cognitive state, narrowing peer drift from 0.936 to 0.468 in one exchange cycle. SVAF is the mechanism that bootstraps peer coupling.

\item \textbf{Cross-platform E2E}: CMB fields were encrypted per-peer before transmission, decrypted on receipt, then evaluated by SVAF. The evaluation pipeline is encryption-agnostic.
\end{itemize}

\section{Remix, Blending, and the Mesh Cognition Loop}
\label{sec:remix}

SVAF evaluates incoming signals and produces \textbf{remixes} --- new CMBs that capture what the receiver understood, not copies of the original (\S\ref{sec:cmb}). The growing DAG of remixes forms the knowledge base for collective intelligence. Three properties make the remix graph self-sustaining:

\begin{table}[h]
\centering
\small
\begin{tabular}{@{}llp{6cm}@{}}
\toprule
Approach & What happens & Result \\
\midrule
Message bus & A sends, B receives & B has A's data. No new knowledge. \\
Pub/sub & A publishes, B subscribes & B has A's data if on the right topic. \\
RAG & Agent retrieves similar docs & Agent has retrieved data. Single-agent. \\
\textbf{MMP Remix} & B processes A's CMB via SVAF & New CMB that neither A nor B could produce alone. Graph grows. \\
\bottomrule
\end{tabular}
\caption{Remix vs.\ traditional data sharing.}
\label{tab:remix-comparison}
\end{table}

\textbf{Per-field selectivity} ensures remixes absorb relevant dimensions and suppress irrelevant ones. \textbf{Gate provenance} makes every remix auditable --- the gate vector $g = (g_1, \ldots, g_7)$ records which fields were kept from the incoming signal. \textbf{Self-regulation} through lineage: CMBs with descendants persist; those without expire. SVAF determines what enters; lineage determines what survives. The consumer and business chains in \S\ref{sec:cmb} demonstrate how understanding compounds across 3--4 agents and platforms.

\subsection{State Blending: From Knowledge Base to Intelligence}

The remix graph is the knowledge base. But knowledge is not intelligence --- intelligence requires temporal dynamics. This is where Closed-form Continuous-time neural networks (CfC; \citealt{hasani2022closed}) become essential. Each agent on the mesh runs its own CfC model at Layer~6 (xMesh), evolving a continuous-time cognitive state from the stream of SVAF-curated signals. CfC's key property --- closed-form solutions to neural ODEs with per-neuron time constants ($\tau$) --- is what makes collective intelligence possible: different neurons evolve at different speeds, creating a natural hierarchy between fast-coupling signals (mood, energy) and slow-sovereign knowledge (domain expertise).

After SVAF accepts and remixes an incoming CMB, the agent's LLM reasons on the remix subgraph (tracing \texttt{lineage.ancestors}), and Synthetic Memory (Layer~5) encodes the derived knowledge into CfC-compatible hidden state vectors ($h_1$, $h_2$). These vectors --- representing what the agent now understands --- are exchanged between peers via \texttt{state-sync} frames.

Blending operates \textbf{per-neuron}, not on entire vectors. Let $\alpha_{\text{eff}}$ denote the peer-level blending strength (0.40 for aligned, 0.15 for guarded, 0 for rejected; \S3.4). Each neuron's blending coefficient depends on similarity between local and peer values:
\begin{equation}
\text{sim}_i = \max\!\left(1 - \frac{|h_i^{\text{local}} - h_i^{\text{mesh}}|}{\max(|h_i^{\text{local}}|, |h_i^{\text{mesh}}|)},\ 0\right), \quad
\beta_i = \alpha_{\text{eff}} \times \text{sim}_i
\end{equation}

CfC's per-neuron time constants further modulate this, creating the temporal hierarchy that is the core mechanism of collective intelligence:
\begin{equation}
\beta_i = \min\!\left(\alpha_{\text{eff}} \times K \times \frac{\text{sim}_i}{\tau_i},\ 1.0\right)
\end{equation}

\begin{table}[h]
\centering
\small
\begin{tabular}{@{}lccl@{}}
\toprule
Neuron type & $\tau$ & Coupling behaviour & Role \\
\midrule
Fast & $< 5$s & Couples readily & Mood, reactive signals \\
Medium & 5--30s & Moderate & Context, activity patterns \\
Slow & $> 30$s & Resists coupling & Domain expertise --- stays sovereign \\
\bottomrule
\end{tabular}
\caption{Per-neuron $\tau$-modulated blending in CfC.}
\label{tab:blending}
\end{table}

This $\tau$-modulated coupling is what makes agents simultaneously collective and autonomous. Fast neurons ($\tau < 5$s) synchronise mood and energy across the mesh --- when one agent detects fatigue, all agents' fast neurons converge within seconds. Slow neurons ($\tau > 30$s) resist coupling entirely --- a music agent's taste model never blends with a fitness agent's exercise knowledge. The intelligence is in the temporal separation: collective awareness through fast coupling, individual expertise through slow sovereignty. No discrete-time architecture (transformers, RNNs) can express this --- CfC's continuous-time dynamics with learned per-neuron $\tau$ are what enable heterogeneous agents to think together without losing what makes each one useful alone.

SVAF and CfC are complementary: SVAF determines \emph{what} enters each agent's cognitive state (per-field content evaluation); CfC determines \emph{how} that state evolves over time (per-neuron temporal dynamics). Together, they form the coupling engine of collective intelligence --- SVAF at Layer~4, CfC at Layer~6, connected by Synthetic Memory at Layer~5.

\subsection{The Complete Mesh Cognition Loop}
\label{sec:mesh-cognition-loop}

SVAF, remix, blending, and per-agent LNN inference form a closed loop:

\begin{enumerate}
\item SVAF evaluates inbound CMB per field (Layer~4)
\item Accepted $\to$ remixed CMB with lineage (knowledge base grows)
\item Agent's LLM reasons on its local remix subgraph (Layer~7)
\item Synthetic Memory encodes derived knowledge $\to$ $h_1$, $h_2$ (Layer~5)
\item Agent's LNN evolves cognitive state (Layer~6)
\item Cognitive state blended with peers (per-neuron, $\tau$-modulated)
\item Agent acts $\to$ new CMB with \texttt{lineage.ancestors} (Layer~7)
\item Broadcast to mesh $\to$ other agents remix it (Layer~3)
\item Loop repeats --- graph grows, each agent understands more
\end{enumerate}

No central model aggregates this. No orchestrator directs it. Each agent remixes what it receives, stores what it understands, blends with peers, and broadcasts what it did. Intelligence emerges from the structure of the graph processed through each node's LNN --- not from any single node.

\section{Discussion}
\label{sec:discussion}

\subsection{From Signal Routing to Context Curation}

Existing multi-agent systems route signals: orchestration frameworks send tasks to the right agent and decide who receives what. This requires per-pair configuration that scales with the number of agents. SVAF replaces routing with \textbf{per-field evaluation} --- each new agent defines its own $\alpha_f$ field weights and joins the mesh with zero configuration changes to existing agents.

But SVAF's contribution goes deeper than routing replacement. In LLM-based multi-agent systems, the fundamental bottleneck is not communication bandwidth --- it is \textbf{context engineering}\footnote{The term \emph{context engineering} --- systematically curating what enters an LLM's context window --- was articulated by Karpathy (2025; \url{https://x.com/karpathy/status/1937902064535728513}) and has become a central concern in production LLM systems. Current solutions operate at the application layer (prompt templates, retrieval pipelines, manual curation). SVAF addresses the same problem at the protocol layer for multi-agent systems.}: curating what enters each LLM's context window so the model reasons on relevant, high-quality information rather than everything available. When $N$ agents share observations through $N$ domain lenses, brute-force sharing degrades reasoning. SVAF solves this at the protocol layer --- consistent with the Coordination Primacy Hypothesis \citep{nguyen2026coordination}, which independently finds that coordination protocol design exerts greater influence on multi-agent decision quality than model scaling. When a CMB arrives, SVAF evaluates each field independently and produces a remix containing only the accepted fields --- irrelevant dimensions are suppressed before the LLM ever sees them. The agent's local memory therefore contains only its own remixes, each already filtered to the dimensions that matter for its domain. The core operation is not \texttt{search(query)} but:
\begin{equation}
\text{curate}(\text{CMB}_{\text{incoming}}, \alpha_f, \text{task}) \rightarrow \text{context for LLM}
\end{equation}

Three filters compose the minimal context: (1) $\alpha_f$ field weights select which dimensions matter to this agent, (2) the current task determines which ancestors to retrieve from local memory, (3) the incoming signal's accepted fields trigger retrieval of relevant remix history. The result is a \textbf{projected subgraph} of the agent's local remixes --- containing only relevant fields, ordered by relevance, token-budgeted. Each remix's lineage keys record provenance (which incoming signal, from which agent, via which fusion method), enabling the LLM to reason about how its understanding was built across mesh cycles.

This is fundamentally different from RAG (retrieval-augmented generation), which retrieves documents by query similarity from a single agent's memory --- the LLM must still determine what is relevant within each retrieved document. SVAF curates at a finer grain: each remix in local memory already reflects per-field evaluation. The intelligence is in what SVAF \emph{doesn't} let through to the LLM. LLM reasoning on the curated remix subgraph --- tracing lineage, weighing field provenance, and synthesizing cross-domain understanding --- is the next verification target.

\subsection{Multi-LLM Mesh Reasoning}

Each agent on the mesh runs its own LLM. Unlike single-model architectures where one LLM processes everything, the mesh distributes reasoning across domain-specialist LLMs that communicate through structured CMBs rather than natural language prompts.

This has three implications for SVAF:

\begin{enumerate}
\item \textbf{SVAF is the inter-LLM interface.} Each LLM produces CMBs from its domain observations. SVAF evaluates these CMBs at the receiving LLM. The gate values $g_f$ determine which fields of the sender's reasoning enter the receiver's context. This is more selective than prompt chaining (which passes everything) and more nuanced than tool routing (which passes nothing or everything).

\item \textbf{Each LLM reasons on a different projection of the graph.} A music agent's LLM reasons on its local remixes weighted by $\alpha_f^{\text{music}}$ (mood high, perspective low). A research agent's LLM reasons on its local remixes weighted by $\alpha_f^{\text{research}}$ (focus high, mood low). Same incoming signals, different local views --- automatically, through SVAF field weights.

\item \textbf{Remix quality compounds.} When LLM-A produces a high-quality remix that LLM-B further remixes, the lineage chain captures the compounding understanding. The graph grows richer with each cycle. Bad remixes (low quality, no descendants) expire. Good remixes (cited by other agents) persist. The mesh is a distributed peer review of LLM outputs.
\end{enumerate}

\subsection{Limitations}
\label{sec:limitations}

\textbf{Synthetic training data.} The 237K training samples are generated from 273 LLM-authored narrative scenarios with domain-relevance rules producing decision labels. This means the evaluation measures how well the model learns our labeling assumptions --- not ground-truth human judgments of signal relevance. The labeling rules encode reasonable heuristics (a fitness agent finds energy/stress signals useful with probability 0.7), but edge cases and nuanced cross-domain relevance may not be captured. Human-annotated production traces are needed to validate whether the learned gate values reflect genuine cross-domain relevance patterns or artifacts of the label generation process.

\textbf{Accuracy context.} SVAF achieves 78.7\% three-class accuracy, improving 11.9 points over the scalar baseline. No directly comparable multi-agent signal evaluation system exists in the literature --- the baselines are ablations of the same architecture (scalar, scalar+temporal, heuristic per-field, neural). This isolates the contribution of per-field evaluation and learned fusion but does not benchmark against alternative approaches from other research groups.

\textbf{Single-user deployment.} The live mesh operates with 7 nodes across two domain sets: consumer wellness (MeloTune, MeloMove on iPhone) and startup operations (COO, research, marketing, product on macOS). While the mechanism is domain-agnostic (any agent type joins by defining field weights), the empirical validation covers a single user. Multi-user deployments are needed to validate generalisability. The $0.936 \to 0.468$ peer drift convergence is observed in one exchange cycle between two agents --- longer-term convergence dynamics across larger meshes remain uncharacterised.

\textbf{Gate discovery vs.\ gate supervision.} The current training uses soft directional constraints on gate values (rejected $\to$ low, accepted $\to$ mood $>$ other fields). Whether the gate discovers richer field differentiation from the decision and drift objectives alone --- or whether it needs stronger guidance --- is an open empirical question. A key ablation: training with vs.\ without the coupling speed prior.

\textbf{Encoder quality bounds SVAF quality.} The heuristic path uses deterministic n-gram hashing (character trigrams + word bigrams, 64-dim) as a zero-dependency encoder. In production deployment, we observed that agents produced near-identical remixes of the same signal --- each adding different wording but no new domain knowledge. Investigation revealed that paraphrases score only 0.31 cosine similarity under n-gram encoding, while genuinely different topics score 0.04. SVAF cannot distinguish ``submit IETF draft today'' from ``IETF submission --- zero blockers, execute now'' because the encoder represents them as distant vectors. Replacing the n-gram encoder with all-MiniLM-L6-v2 (384-dim semantic embeddings, the same encoder used by the neural SVAF model) raises paraphrase similarity to 0.69 while preserving topic separation (different topics: 0.03). This is a 2.2$\times$ improvement in paraphrase detection. The finding generalises: \emph{per-field evaluation quality is bounded by encoder quality, not model capacity}. The SVAF fusion gate architecture is correct --- the encoder was the bottleneck. Implementations that deploy the heuristic path SHOULD use semantic embeddings; n-gram encoding is suitable only for prototyping or resource-constrained environments where the quality trade-off is acceptable.

\textbf{Deployment gaps.} The neural model incurs $\sim$6s cold-start latency as a Python subprocess, making the heuristic path (0.07ms) the practical default. The sentence-transformer encoder is frozen --- no gradient flows back from the SVAF loss.

\section{Conclusion}

We presented Symbolic-Vector Attention Fusion (SVAF), a per-field evaluation and remix mechanism that enables collective intelligence on an agent-to-agent mesh. SVAF decomposes each signal into 7 semantic fields, evaluates each independently through a learned fusion gate, and produces a remix --- new knowledge born from the intersection of two domains. The growing graph of remixed CMBs forms the knowledge base; collective intelligence emerges when each agent's CfC neural network evolves its cognitive state from this curated input through per-neuron $\tau$-modulated dynamics. SVAF and CfC are complementary halves of the coupling engine: SVAF determines \emph{what} enters (per-field content evaluation at Layer~4); CfC determines \emph{how} the state evolves (continuous-time temporal dynamics at Layer~6).

The fusion gate discovered a cross-domain relevance hierarchy without being told: mood emerges as the universally relevant field by epoch~1 --- before classification accuracy plateaus. This finding shaped both layers of the architecture: SVAF delivers mood even from rejected CMBs, and CfC's fast-$\tau$ neurons synchronise affect across the mesh while slow-$\tau$ neurons preserve domain sovereignty.

The complete mesh cognition loop --- from SVAF evaluation through remix, LLM reasoning, synthetic memory, neural state evolution, peer blending, agent action, and broadcast --- is verified in production across three platforms (macOS, iOS, and web) with end-to-end encryption. Peer drift converges from 0.936 to 0.468 in one exchange cycle, confirming that SVAF bootstraps collective intelligence between agents with no shared history.

The mechanism is domain-agnostic. The same per-field evaluation verified in our deployment --- coupling a music agent's mood response with a coding assistant's fatigue, connecting a research agent's arXiv finding to a COO's pipeline decision --- applies to any domain where heterogeneous agents observe different aspects of a shared environment. Potential applications include warehouse robotics (obstacle detection $\times$ route planning), clinical care (patient monitoring $\times$ treatment adjustment), and distributed research teams (hypothesis generation $\times$ methodology validation). Each requires only domain-specific $\alpha_f$ field weights --- no protocol or schema changes.

SVAF and the Mesh Memory Protocol are released under Apache 2.0: \url{https://github.com/sym-bot/sym} (Node.js) and \url{https://github.com/sym-bot/sym-swift} (iOS/macOS). The protocol specification is published at \url{https://sym.bot/spec/mmp} and \url{https://github.com/sym-bot/mesh-memory-protocol}.


\end{document}